\documentclass{article}
\usepackage{arxiv}
\usepackage[utf8]{inputenc} 
\usepackage{graphicx}
\usepackage{subcaption}
\usepackage[T1]{fontenc}    
\usepackage{hyperref}       
\usepackage{url}            
\usepackage{booktabs}       
\usepackage{amsfonts}       
\usepackage{nicefrac}       
\usepackage{microtype}      
\usepackage{lipsum}
\usepackage{xcolor}
\begin{document}

\title{Generalised thresholding of hidden variable network models with scale-free property}
\author{
  S\'{a}muel G. Balogh\\
  Dept. of Biological Physics\\ 
  E\"{o}tv\"{o}s University\\
  H-1117 Budapest, Hungary\\
  \texttt{balogh@hal.elte.hu} \\
   \And
 P\'{e}ter Pollner \\
  MTA-ELTE Statistical and Biological Physics Research Group\\
  Hungarian Academy of Sciences\\
  H-1117 Budapest, Hungary\\
   \AND
   Gergely Palla \\
  MTA-ELTE Statistical and Biological Physics Research Group\\
  Hungarian Academy of Sciences\\
  H-1117 Budapest, Hungary\\
}


\begin{abstract}
The hidden variable formalism (based on the assumption of some intrinsic node parameters) turned out to be a remarkably efficient and powerful approach in describing and analyzing the topology of complex networks. Owing to one of its most advantageous property - namely proven to be able to reproduce a wide range of different degree distribution forms - it has become a standard tool for generating networks having the scale-free property. One of the most intensively studied version of this model is based on  a thresholding mechanism  of  the  exponentially distributed  hidden variables associated to the nodes (intrinsic vertex weights), which give rise to  the emergence of a scale-free network where the degree distribution $p(k)\sim k^{-\gamma}$ is decaying with an exponent of $\gamma =2$. Here we propose a generalization and modification of this model by extending the set of connection probabilities and hidden variable distributions that lead to the aforementioned degree distribution, and analyze the conditions leading to the above behavior analytically. In addition, we propose a relaxation of the hard threshold in the connection probabilities, which opens up the possibility for obtaining sparse scale free networks with arbitrary scaling exponent.
\end{abstract}

\flushbottom
\maketitle

\thispagestyle{empty}

\section*{Introduction}

Network theory provides an ubiquitous and sophisticated approach for the characterization of complex systems, possibly composed of many interacting units \cite{Laci_revmod,Dorog_book}. One of the most widely studied features of complex networks is given by the scale-free (SF) property, manifesting in the strong inhomogeneity of the degree distribution, $p(k)$, accompanied by a power-law like decay of $p(k)$ in the large degree regime \cite{barabasi,newmancollaborationnet,sfbrainactivity,caldarelli_general_sf_recipe,albertreka}. On the modelling ground, several growing mechanisms have been proposed for generating networks without a characteristic degree scale, including the fundamental concept of the Barab\'{a}si-Albert model together with its modifications and generalizations \cite{barabasi,bianconibarabasi,dorogpreflink}.
Meanwhile it became also evident that not all networks emerge from a growing mechanism \cite{caldarelli_first_peaked,sfexponone,sfexponlessthantwo}, and there are numerous examples where new connections can easily occur between already existing nodes in the system \cite{Watts-Strogatz,growing_with_intrinsic_vertex_fit}. In some cases we can also reasonably assume that the rewiring of the network leads towards a more optimal configuration \cite{Roger_optimal,our_graph_ensemble,scirep_optimal}, and that the propensity for creating links is encoded in each node as an intrinsic parameter. Assumptions of this type naturally gave rise to the development of the hidden variable formalism.

Inspired by the nature of protein interaction networks, the hidden variable model was originally introduced by Caldarelli {\it et al.\ }for explaining the emergence of non-growing scale-free networks, where link creation might be related to some intrinsic features of the nodes \cite{caldarelli_first_peaked}. In a following work, Bogu\~na {\it et al.\ }proposed a systematic analytic framework for generally characterizing classes of random graphs generated with hidden variables \cite{boguna_general_hv}. Based on this framework, later on a general method was implemented, capable of producing SF networks with a tunable $\gamma$ scaling exponent \cite{caldarelli_general_sf_recipe}. Since then, the applicability of the hidden variable formalism has been confirmed on a large scale \cite{vipclub,local,masuda_inverse_square,fujihara_extreme_theory_hv,edge_picking_hv,hidden_social,satorras_activity_driven_hv_mapping,bak_sneppen}, and due to its very general nature, a large variety of further network models ranging from stochastic block models \cite{Holland_SBM,Fienberg_Meyer_Wasserman_SBM,Moore_SBM,Clauset_SBM,Tiago_SBM} through multifractal graph generators \cite{pallalovasz} {\color{black} and evolving, fitness based network models combined with preferential attachment \cite{citation_dynamics,temp_eff_in_growing} to networks defined over hidden metric spaces \cite{Serrano_hidden_metric,curvaturetemperaturecomplexnetw,Boguna_nature_phys,hypnetworks,Boguna_nat_comms}} can be alternatively interpreted as special forms of this approach.

Although many different variations of this model has been proposed, the basic idea of the concept is to first associate a parameter (hidden variable) to the nodes, usually drawn from a prescribed distribution, and then connect the pairs of nodes according to a probability given by a fixed linking function taking the node variables as arguments. A part of these models can be referred to as geographical, where besides the hidden variables, node also have coordinates in a $d$ dimensional Euclidean space, and the connection function is depending on both the distance and the hidden variables \cite{geographthreshgraph}. A very closely related class of models is where the coordinates are distributed in a hyperbolic space instead of a Euclidean one \cite{hypnetworks,curvaturetemperaturecomplexnetw}, which provide a very interesting direction for research on their own, due to that they can generate SF networks with a high clustering coefficient in a natural way and due to their relevance in routing problems \cite{hypnetworks}.

A widely studied version of the hidden variable approach is where the linking function acts as a threshold, giving a connection probability 1 when the value of the two variables fulfill some criteria, and 0 otherwise. Non-geographical threshold models  of this kind have gained considerable attention \cite{hypnetworks,caldarelli_first_peaked,masuda_inverse_square,fujihara_extreme_theory_hv,rigorous_results_on_threshold_netw,limit_thres,Yusuke_threshold_2010}, later on being extended even to the geographical space \cite{geographthreshgraph}. An interesting phenomenon observed in the non-geographical thresholded model is that in the infinitely large system size limit the degree distribution seems to be universally characterized by a $\propto k^{-2}$ decay for various fitness distributions \cite{masuda_inverse_square,fujihara_extreme_theory_hv}. In the present paper we extend the previously studied families of hidden variable distributions and linking functions that fall into this class, and study the mathematical conditions leading to this specific degree decay exponent  analytically. In addition, we also introduce a simple and intuitive relaxation of the former 'hard' threshold functions, that allows the modification of the exponent according to numerical simulations.

\section*{The hidden variable model}
When generating a network with $N$ nodes in this approach, first we need to assign a hidden variable  $\left \{ x_i \ | \ x_i \geq 0 \ \forall i \in \{1,\dots,N\}\right \}$ to each node $i$, where $x_i$ is drawn from an arbitrary (but normalized) $\rho(x)$ probability distribution. For simplicity, $x_i$ is often referred to as the fitness of node $i$. After distributing these intrinsic parameters we also have to define a linking function $0\leq f(x,y)\leq 1$, based on which the connection probability between nodes $i$ and $j$ can be simply expressed as $p($link~between~$i$~and~$j)=f(x_i,x_j)$. Thus, in this model all of the information and the properties of the emerging network is completely encoded in the pre-defined form of $\rho(x)$ and $f(x,y)$.

Following the continuous approximation introduced in \cite{caldarelli_general_sf_recipe,masuda_inverse_square}, the expected degree of nodes with fitness $x$ can be expressed as
\begin{equation}
k(x)=N\int\limits_{y_{min}}^{y_{max}}f(x,y) \rho(y) \mathrm{d}y
\label{eq:conti}
\end{equation}
Assuming that $k(x)$ is a monotonically increasing and invertible function of $x$, according to the rule of transformation of random variables \cite{caldarelli_general_sf_recipe} the degree distribution can be written as 
\begin{equation}
p(k)=\int \rho(x) \ \delta\left(k-k\left(x\right)\right)\mathrm{d}x =\left.\rho(x)\frac{\mathrm{d} x}{\mathrm{d} k(x)} \right|_{\tilde{x}(k)},
\label{eq:av_deg_formula}
\end{equation}
where $\tilde{x}(k)$ denotes the inverse function of $k(x)$.

The simplest choice for the connection function is $f(x,y)=$ const., where the connection probability is uniform and independent from the hidden variables, leading to the emergence of an Erd{\H o}s-R{\'e}nyi random graph. A more interesting example was shown in \cite{caldarelli_first_peaked} with a linking function $f(x,y)=\frac{xy}{x^2_M}$ and $x_M=\max_{i}\{x_i| \ i=1,...,N\}$ where (apart from multiplicative constants) the degree distribution inherits the form of the fitness distribution $p(k)\sim \rho \left(\frac{x^2_{M}}{N\left< x\right >}k\right)$. Thus, by choosing a fitness distribution of $\rho(x) \sim x^{-\gamma}$, the obtained network certainly displays the SF property with the same $\gamma$ exponent. 

Later on it turned out that for a more general class of linking functions where $f(x,y)$ can be  decomposed into a product such as $f(x,y)=g(x)g(y)$, the hidden variable formalism can be regarded as analytically well-treatable, and it is also able to reproduce fat-tailed degree distributions with arbitrary $\gamma$ exponents \cite{caldarelli_general_sf_recipe}. The product form also implies that for randomly chosen links the degrees of the endpoints are un-correlated, thus, from the point of view of degree assortativity the obtained networks are neutral.

Another surprising result in Ref.\cite{caldarelli_first_peaked} is connected to the case where an exponential fitness distribution $\rho(x)\sim \mathrm{e}^{-x}$ is chosen together with an $f(x,y)$ corresponding to a threshold function
\begin{equation}
f(x,y)=\Theta(x+y-\Delta),
\label{eq:nongeothres}
\end{equation}
where $\Theta(x)$ refers to the Heaviside step function, and $\Delta$ is a constant with a logarithmic dependency on the system size $N$. Under these settings a power law decay of the degree distribution was detected with a $\gamma=2$ scaling exponent, providing the first evidence for that SF networks can be generated in this approach even with non power-law like fitness distributions. As we already mentioned in the Introduction, in later studies it was observed that the inverse square decay of the degree distribution is actually a quite general feature,
that holds for various other fitness distributions as well \cite{masuda_inverse_square,fujihara_extreme_theory_hv}. Further interesting occurrences of the inverse square decay is briefly discussed in \cite{calda_minus_2_real_networks}.

\section*{Generalized classes of non-geographical thresholded SF hidden variable models with $\gamma=2$}

\subsection*{Model class definitions}
\label{sect:model_class_def}

In this section we introduce a broad set of hidden variable models where the degree decay exponent is equal to $\gamma=2$. A common feature of these models is that they are thresholded in the sense that the linking function $f(x,y)$ has a lower cut-off, controlled by a parameter $\Delta$. The example in (\ref{eq:nongeothres}) is a special case of this, where  $f(x,y)$ immediately becomes 1 above the threshold. Here we use a much weaker assumption, namely that $f(x,y)$ is 0 for a certain range of $x$ and $y$ values, and is non-zero (but not necessarily 1) outside this range.  This is a far more general way of thresholding the connection probabilities, which allows a very broad range of connection functions to be used in the model, as shall be shown later. The linking functions having this property are denoted as $f_{\Delta}(x,y)$ throughout the paper.
Here we introduce sub-classes of hidden variable models, the first one is to which we refer as exponential-like and where 
\begin{itemize}
\item  $\rho(x)$ can be written as 
\begin{equation}
\rho(x)=H'(x)\exp\left[-H(x)\right],
\label{eq:genrhoexp}
\end{equation} 
where $H(x)$ is a differentiable, monotonously increasing function and $H'(x)$ denotes its derivative,
\item while the thresholded $f_{\Delta}(x,y)$ shows an additive dependency on $H(x)$ and $H(y)$, 
\begin{equation}
f_{\Delta}(x,y)=\left\{
\begin{array}{ll}
0,     & \mbox{if } H(x)+H(y) \leq \Delta,  \\
\tilde{f}(H(x)+H(y))\in [0,1],     & \mbox{if } H(x)+H(y) > \Delta,
\end{array}\right.
\label{eq:genfdeltaexp}
\end{equation}
where $\tilde{f}(x,y)$ is assumed to be a general function taking values in the $[0,1]$ interval.
\end{itemize}
We refer to the second sub-class as power-like, where
\begin{itemize}
\item $\rho(x)$ can be written as
\begin{equation}
\rho(x)=G'(x)G^{-\alpha}(x), 
\label{eq:genrhopow}
\end{equation}
where $G(x)$ has the same properties as $H(x)$ in (\ref{eq:genrhoexp}),
\item while the thresholded $f_{\Delta}(x,y)$ shows a multiplicative dependency on $G(x)$ and $G(y)$,
\begin{equation}
f_{\Delta}(x,y)=\left\{
\begin{array}{ll}
0,     & \mbox{if } G(x)G(y)\leq \Delta,  \\
\tilde{f}(G(x)G(y))\in[0,1]     & \mbox{if } G(x)G(y) >\Delta.
\end{array}\right.
\label{eq:genfdeltapow}
\end{equation}
\end{itemize}

And last, if both additive and multiplicative dependency are present (mixed class): 
\begin{itemize}
\item $\rho(x)$ can be written as
\begin{equation}
\rho(x)=\frac{M'(x)}{\left(1+M(x)\right)^{\alpha}}, 
\label{eq:genrholom}
\end{equation}
where $M(x)$ has the same properties as $H(x)$ in (\ref{eq:genrhoexp}),
\item while $f_{\Delta}(x,y)$ can be expressed as:
\begin{equation}
f_{\Delta}(x,y)=\left\{
\begin{array}{ll}
0,     & \mbox{if } (1+M(x))(1+M(y))\leq \Delta,  \\
\tilde{f}(1+M(x)+M(y)+M(x)M(y))\in[0,1]     & \mbox{if } (1+M(x))(1+M(y)) >\Delta.
\end{array}\right.
\label{eq:genfdeltalom}
\end{equation}
\end{itemize}

These generalized sub-classes can be derived from two simple observations. First, it can be shown in general that when replacing the step-like function in (\ref{eq:nongeothres}) by an arbitrary $f_{\Delta}(x+y)$ having a lower cut-off at $\Delta$, the degree distribution of the emerging networks is not affected. Second, the form of the hidden variable distributions and accompanying connection functions given in (\ref{eq:genrhoexp}-\ref{eq:genfdeltalom}) are also in very close relation with the transformations of the $(\rho,f)$ pair that leave the degree distribution of the generated network invariant. To see that, let us assume an arbitrary $\rho(x)$ and $f(x,y)$ yielding a network with a degree distribution of $p(k)$. By transforming the hidden variables using a monotonous function $H$ as $x_i=H(z_i)$ and $z_i = H^{-1}(x_i)$  for all nodes $i$, according to the rule of transforming random variables the density of the original variable $x$ can be also written as $\rho_x(x)=\rho_z(z)/H'(z)=\rho_z(H^{-1}(x))/H'(H^{-1}(x))$. Based on that, the expected degree for nodes with variable $x$ given in (\ref{eq:av_deg_formula}) can be also expressed as
\begin{equation}
    k(x) = k(H(z)) = N\int\limits_0^{\infty} f(x,y)\rho_x(y)\mathrm{d}y = 
    N\int\limits_0^{\infty}f(H(z),H(z'))\rho_z(z')\mathrm{d}z',
    \label{eq:symm}
\end{equation}
where we have changed the integration variable $y$ to $z'=H^{-1}(y)$. By combining the expression of (\ref{eq:symm}) with the transformation rule of degrees given in  (\ref{eq:av_deg_formula}), we obtain that a transformed model where $\rho_z(z)=\rho_x(H(z))H'(z)$ and the linking function is given by $f(H(x),H(y))$, will essentially lead to the same degree distribution as the original model. This conservation law of the degree distribution is also closely related to a general
transformation rule of the fitness values written as
\begin{equation}
    p(k)=
    \left. \frac{\mathrm{d}H(z)}{\mathrm{d}z} \rho\left(H(z)\right) \frac{\mathrm{d} x}{\mathrm{d} \int\limits_{0}^{\infty}f\left(H(z),H(z')\right)\frac{\mathrm{d}H(z')}{\mathrm{d}z'}\rho\left(H(z')\right)\mathrm{d}z'}\right|_{\tilde{z}_{H}(k)},
\end{equation}
which is analogous to (\ref{eq:av_deg_formula}).
The sub-classes of models we defined in this paper exploit this property, where the 'original' model is corresponding to the simple model introduced in \cite{caldarelli_first_peaked},  following the inverse square decay law. Nevertheless, this invariance of the degree distribution under appropriate simultaneous transformation of $\rho(x)$ and $f(x,y)$ is valid for the hidden variable approach in general, also in the case of geographical models. 
We also note that any model in one of the above defined three sub-classes can generally be mapped into a model in the other sub-class via simple transformations between $H(x)$, $G(x)$ and $M(x)$ given by 
 \begin{equation}
 H(x)=\alpha\ln G(\tilde{x}) ,\ \ \ \  H(x)=\alpha\ln \left(1+M(\tilde{x})\right) ,\ \ \ \  G(x)=1+M(\tilde{x}).     \label{eq:transformation}
 \end{equation}
However, when the goal is to obtain a size independent SF degree distribution, the dependency of $\Delta=\Delta(N)$ on the number of nodes $N$ can be different in each class.
 In summary, the previous observations clarify in a simple manner why seemingly different realizations of $(\rho,f)$ can give rise to the emergence of networks with the same degree distributions. In addition, we also gained simple rules for mapping the different realisations of $(\rho,f)$ into each other. A remarkable consequence of the above is that for any $\tilde{f}_{\Delta}$ showing either additive, multiplicative or mixed dependency on its arguments, we can now construct a fitness distribution for which it is guaranteed that the degree distribution of the  emerging network will display the $p(k)\sim k^{-2}$ behaviour.

\begin{figure}[htb!]
  \begin{center}
  \centering
    \includegraphics[width=0.5\textwidth]{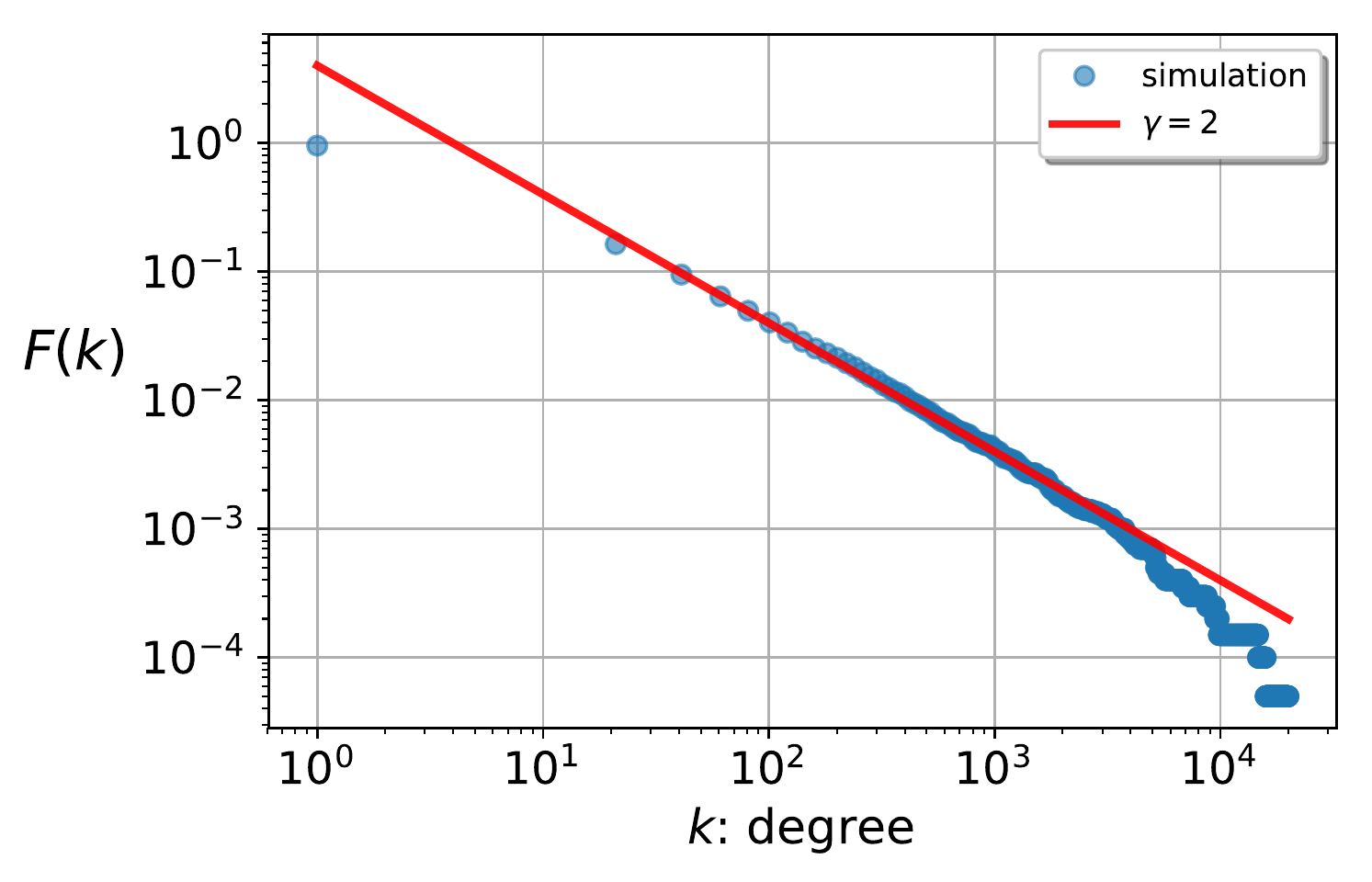}
      \caption{Complementary  cumulative distribution of the node degrees $F(k)$ in a network of size $N=20000$, obtained from simulations for the Weibull fitness distribution in Table.\ref{tab:partforms}, at a scale parameter $c=2$, and a linking function defined in (\ref{eq:Weibul_f}), shown on logarithmic scale. The solid line is decreasing as $k^{-1}$, which is corresponding to the decay characteristics of $F(k)$ in SF networks with $\gamma=2$.} 
      \label{fig:c_finite}
  \end{center}
\end{figure}

For illustration, a few examples from each sub-classes are listed in Table.\ref{tab:partforms}, all generating SF networks with a degree decay exponent $\gamma=2$. In addition, in Fig.\ref{fig:c_finite}. we show the simulation results for the Weibull fitness distribution from the class of exponential-like distributions, where the corresponding linking function was chosen as
\begin{equation}
f_{\Delta}(x,y)=
\left\{ \begin{array}{ll}
0 & \mbox{\rm if } x^2+y^2 \leq \Delta, \\
1-\frac{\exp\left[a-\left(x^2+y^2\right)\right]}{\left(x^2+y^2\right)} 
& \mbox{if } x^2+y^2 > \Delta,
\end{array} \right.
\label{eq:Weibul_f}
\end{equation}
 where $a$ is a constant and $\Delta$ is defined via the transcendent equation $\Delta=\exp(a-\Delta)$. The fact that the complementary cumulative distribution $F(k)=\int_{k'=k}^{\infty}p(k')\mathrm{d}k'$ of the degrees behaves as $k^{-1}$ in the large degree regime in Fig.\ref{fig:c_finite}. is in full consistency with the inverse square decay law.
 
\begin{table}[htb!]
\begin{center}
 \begin{tabular}{||c c c||}
 \hline \\[-1em]
 \rule{0pt}{15pt} $\rho(x)$  & Name of the distribution &  $f_{\Delta}(x,y)$ \\ [1.0ex]
 \hline \\[-1em]
 \rule{0pt}{15pt} & \textbf{\textit{Exponential-like $\rho(x)$,}} &   \\ [1.0ex]
  & \textbf{\textit{additive $f_{\Delta}(x,y)$}} & \\
 \rule{0pt}{15pt}  $\mathrm{e}^{-x}$ & exponential &  $f_{\Delta}(x+y)$ \\ [1ex]
 
 \rule{0pt}{15pt}  $cx^{c-1}\mathrm{e}^{-x^c}$ & Weibull &  $f_{\Delta}\left(x^c+y^c\right)$ \\ [1ex]
 
 \rule{0pt}{15pt} $b\eta \mathrm{e}^{bx}\exp\left(-\eta \mathrm{e}^{bx}\right)$ & Gompertz &  $f_{\Delta}\left(\eta \mathrm{e}^{bx}+\eta \mathrm{e}^{by}\right)$ \\ [1ex]
 \hline
 \rule{0pt}{15pt} & \textbf{\textit{Power-like $\rho(x)$,}} &   \\ [1.0ex]
  & \textbf{\textit{multiplicative $f_{\Delta}(x,y)$}}& \\

  \rule{0pt}{15pt} $Cx^{-\alpha}$ & power &   $f_{\Delta}(xy)$  \\ [1ex]
  \rule{0pt}{15pt}  $cx^{c-1}\mathrm{e}^{-x^c}$ & Weibull &  $f_{\Delta}\left(\exp\left(x^c\right)\exp\left(y^c\right)\right)$ \\ [1ex]
 \hline
 \rule{0pt}{15pt} & \textbf{\textit{Mixed $\rho(x)$,}} &   \\ [1.0ex]
  & \textbf{\textit{Mixed $f_{\Delta}(x,y)$}}& \\

 \rule{0pt}{15pt} $\frac{\mathrm{e}^{\frac{x-\mu}{s}}}{\left(1+\mathrm{e}^{\frac{x-\mu}{s}}\right)^2}$ & logistic &   $f_{\Delta}\left(\left(1+\mathrm{e}^{\frac{x-\mu}{s}}\right) \left(1+\mathrm{e}^{\frac{y-\mu}{s}}\right)\right)$ \\ [1ex]
 \rule{0pt}{15pt}  $\frac{\frac{\eta}{c}\left ( \frac{x}{c} \right )^{\eta-1}}{\left ( 1+\left ( \frac{x}{c} \right )^{\eta} \right )^2}$  & Fisk & $f_{\Delta}\left(\left(1+\left ( \frac{x}{c} \right )^{\eta}\right) \left(1+\left ( \frac{x}{c} \right )^{\eta}\right)\right)$\\ [1ex]
 \hline
\end{tabular}
\end{center}
\caption{Examples for fitness distributions with the accompanying thresholded fitness functions $f_{\Delta}(x,y)$ for each sub-class. The listed models all lead to the emergence of SF networks satisfying the inverse square decay law.}
\label{tab:partforms}
\end{table}

\subsection*{The inverse square decay law}
\label{sect:inverse_square_decay_law}

Here we show in details that for thresholded hidden variable models falling into the class described in the previous subsection, the degree distribution of the emerging SF network will always have a degree decay exponent of $\gamma=2$. Let us assume that we are dealing with an exponential-like model, where the fitness distribution is given in (\ref{eq:genrhoexp}), and the linking function follows (\ref{eq:genfdeltaexp}).
Starting from the expression for the average degree given in (\ref{eq:av_deg_formula}), and multiplying both sides by $\exp\left[-H(x)\right]$ we can write
\begin{equation}
k(x)\exp\left[-H(x)\right]=N\int\limits^{\infty}_{H^{-1}(\Delta-H(x))}\tilde{f}\left[H(x)+H(y)\right]H'(y)\exp\left[-H(x)-H(y)\right]\mathrm{d}y,
\label{eq:general_derivation_first_form}
\end{equation}
where we used that $f_{\Delta}(x,y)=0$ if $H(x)+H(y)\leq \Delta$. By a change of variable $z=H(x)+H(y)$ we arrive to an equation where the right hand side is independent of $x$,
\begin{equation}
k(x)\exp\left[-H(x)\right]=N\int\limits^{\infty}_{\Delta}\tilde{f}(z)\mathrm{e}^{-z}\mathrm{d}z.
\label{eq:change_variable}
\end{equation}
Based on (\ref{eq:change_variable}) we define the integral 
\begin{equation}
L_{\Delta}(\tilde{f}) \equiv \int\limits_{\Delta}^{\infty}\tilde{f}(z)\mathrm{e}^{-z}\mathrm{d}z,
\label{eq:L_def}
\end{equation}
that depends on the form of the actually chosen $\tilde{f}(z)$ appearing in (\ref{eq:genfdeltaexp}). Assuming that this integral exists, using (\ref{eq:change_variable}-\ref{eq:L_def}) we can express the average degree of nodes with fitness $x$ and the derivative of $k(x)$ as
\begin{eqnarray}
k(x)&=&NL_{\Delta}(\tilde{f})\exp\left[H(x)\right], \label{eq:k_x}\\
\frac{\mathrm{d}k(x)}{\mathrm{d}x}&=&NL_{\Delta}(\tilde{f})\exp\left[H(x)\right]H'(x).
\label{eq:dk_x}
\end{eqnarray}
In the thermodynamic limit of $N\rightarrow\infty$, by substituting (\ref{eq:k_x}-\ref{eq:dk_x}) into (\ref{eq:av_deg_formula}) we obtain
\begin{equation}
p(k)=\left.  \frac{H'(x)\exp\left[-H(x)\right]}{NL_{\Delta}(\tilde{f})H'(x)\exp\left[H(x)\right]}\right |_{\tilde{x}_H(k)}=\left.\frac{NL_{\Delta}(\tilde{f})}{\left[NL_{\Delta}(\tilde{f})\exp\left[H(x)\right]\right]^2}\right |_{\tilde{x}_H(k)}=\frac{NL_{\Delta}(\tilde{f})}{k^2},
\label{eq:inversesquare}
\end{equation}
showing that $p(k)$ is proportional to $k^{-2}$, since $L_{\Delta}(\tilde{f})$ is a constant. Moreover, according to (\ref{eq:inversesquare}) we can formulate a very simple condition under which $p(k)$ becomes independent of the system size in the form of 
\begin{equation}
L_{\Delta}(\tilde{f})=L(\Delta,\tilde{f})=\int\limits_{\Delta}^{\infty}\tilde{f}(z)\mathrm{e}^{-z}\mathrm{d}z=\frac{1}{N},
\label{eq:exp_condition}
\end{equation}
which is reducing to $L(\Delta,\Theta)=\mathrm{e}^{-\Delta}=\frac{1}{N}$ in the case of the simple step-function like connection function given in (\ref{eq:nongeothres}), in complete agreement with the results in \cite{caldarelli_general_sf_recipe,boguna_general_hv}. 

Similarly to the case of exponential-like distributions, if we choose the fitness distribution to be power-like as in (\ref{eq:genrhopow}), together with a thresholded connection function given in (\ref{eq:genfdeltapow}), the expected degree for a node having a hidden variable $x$ can be written as
\begin{equation}
k(x)=N\int\limits^{\infty}_{1}f(x,y) \rho(y) \mathrm{d}y=N\int\limits^{\infty}_{G^{-1}(\Delta/G(x))}\tilde{f}\left(G(x)G(y)\right)G^{-\alpha}(y)G'(y)\mathrm{d}y.
\end{equation}
By changing to the integration variable $z=G(x)G(y)$ we obtain
\begin{equation}
k(x)G^{-\alpha+1}(x)=N\int\limits^{\infty}_{\Delta}\tilde{f}\left[G(x)G(y)\right]G^{-\alpha}(x)G^{-\alpha}(y)\mathrm{d}[G(x)G(y)],
\end{equation}
leading to
\begin{equation}
k(x)=NG^{\alpha-1}(x)\int\limits^{\infty}_{\Delta}\tilde{f}(z) z^{-\alpha}\mathrm{d}z,
\label{eq:pow_k_x_sub}
\end{equation}
{\color{black} where $\alpha > 1$.} The integral 
\begin{equation}
 K_{\alpha}(\tilde{f})\equiv \int\limits^{\infty}_{\Delta}\tilde{f}(z) z^{-\alpha} \mathrm{d}z
\label{eq:sizeindep_pow}
\end{equation}
appearing in (\ref{eq:pow_k_x_sub}) depends only on the chosen form of $\tilde{f}(z)$, thus, by assuming that $K_{\alpha}(\tilde{f})$ exists we can express $k(x)$ simply as
\begin{equation}
k(x) = NK_{\alpha}(\tilde{f}) G^{\alpha-1}(x).
\end{equation}
By substituting this into the general formula for the degree distribution given in (\ref{eq:av_deg_formula}) we gain
\begin{equation}
p(k)=\left. \rho(x)\frac{dx}{dk(x)}\right|_{\tilde{x}_G(k)}=\left. \frac{G'(x)G^{-\alpha}(x)}{N(\alpha-1)K_\alpha(\tilde{f})G'(x) G^{\alpha-2}(x)}\right|_{\tilde{x}_G(k)} 
\sim \left. \frac{1}{G^{2(\alpha-1)}(x)}\right|_{\tilde{x}_G(k)}\sim k^{-2},
\end{equation}
proving that the power-like sub-class also leads to the emergence of a SF network with $\gamma=2$. However, in this case the condition for a size independent degree distribution with the $p(k)\sim k^{-2}$ property is given by a different formula compared to (\ref{eq:exp_condition}), written as 
\begin{equation}
 K_{\alpha}(\tilde{f})\equiv \int\limits^{\infty}_{\Delta}\tilde{f}(z) z^{-\alpha} \mathrm{d}z=\frac{1}{N}.
\label{eq:const_sizeindep_pow}
\end{equation}
By following similar mathematical arguments as we did in the exponential- and power-like cases the same behaviour can be obtained for the third sub-class defined through (\ref{eq:genrholom}-\ref{eq:genfdeltalom}). Nevertheless, the condition for a size independent SF degree distribution with $\gamma=2$ has yet again, a different from (\ref{eq:exp_condition}) or (\ref{eq:const_sizeindep_pow}), given by
\begin{equation}
 J_{\alpha}(\tilde{f})\equiv \int\limits^{\infty}_{\Delta}  \frac{\tilde{f}(z)}{(1+z)^{\alpha}} \mathrm{d}z=\frac{1}{N}.
\label{eq:sizeindep_lom}
\end{equation}
An important related remark is that for all fitness distributions and linking functions given in the forms of (\ref{eq:genrhoexp}), (\ref{eq:genfdeltaexp}) or (\ref{eq:genrhopow}), (\ref{eq:genfdeltapow}) or (\ref{eq:genrholom}), (\ref{eq:genfdeltalom}) the emerging networks always display the inverse square decay law independent from the specific form of $\tilde{f}$, and thus, the appropriate form of $\tilde{f}$ only determines the condition for obtaining size-independent degree distribution.

\section*{Soft thresholding with tunable degree decay exponent}

The degree decay exponent of SF complex networks characterizing real systems is usually between $\gamma=2$ and $\gamma=3$ \cite{Laci_revmod,Dorog_book}. Motivated by that, we extend the models defined in the previous section to allow the emergence of hidden variable networks with a $\gamma>2$ exponent as well in the same framework. 

The basic idea is to relax the 'hard' threshold in the connection function, controlled by the variable $\Delta$ in (\ref{eq:nongeothres}) and (\ref{eq:genfdeltaexp}).  
Let us start with the Heaviside step function given in (\ref{eq:nongeothres}), which we can intuitively replace by a 'reversed' Fermi-Dirac function 
\begin{equation}
    f(x,y) = f(x+y) = \frac{1}{1+\mathrm{e}^{-\beta(x+y-\Delta)}},
    \label{eq:soft_thresh_base}
\end{equation}
where $\beta$ is a parameter taking positive values. In the $\beta\rightarrow\infty$ limit we recover the original step-like connection function (\ref{eq:nongeothres}), whereas for finite $\beta$ values we obtain a 'soft' threshold function.The form of $f(x,y)$ in (\ref{eq:soft_thresh_base}) is similar to that of the linking probability in temperature dependent graph ensembles \cite{Garlaschelli_Entropy,origin_of_deg_corrs}. The interpretation of $\beta$ in this respect is analogous to the inverse temperature, with $\beta\rightarrow\infty$ corresponding to the zero temperature 'ground' state, while networks generated with finite $\beta$ values can be interpreted as states at higher temperatures \cite{Garlaschelli_Entropy,curvaturetemperaturecomplexnetw}. %
Regarding the case where we can not assume the additive but rather the multiplicative dependency on $x$ and $y$ in (\ref{eq:soft_thresh_base}), a simple approach to relax the step function is to define $f(x,y)$ as
\begin{equation}
    f(x,y)= f(xy) = \frac{1}{1+\left(\frac{xy}{\Delta}\right)^{-\beta}},
\end{equation}
converging to $f(x,y)=\Theta(xy-\Delta)$ in the $\beta\to \infty$ limit.

In a similar fashion to (\ref{eq:soft_thresh_base}), for the general exponential-like models defined in (\ref{eq:genrhoexp}-\ref{eq:genfdeltaexp}) we can replace (\ref{eq:genfdeltaexp}) by
\begin{equation}
    f_{\beta,\Delta}(x,y)=\frac{1}{1+\mathrm{e}^{-\beta(H(x)+H(y)-\Delta)}},
    \label{eq:soft_thresh_exp}
\end{equation}
and for the power-like models given in (\ref{eq:genrhopow}-\ref{eq:genfdeltapow}) we can change (\ref{eq:genfdeltapow}) to
\begin{equation}
    f(x,y) =  f_{\beta,\Delta}(x,y)=\frac{1}{1+\left(\frac{G^{\alpha}(x)G^{\alpha}(y)}{\Delta}\right)^{-\beta}}.
    \label{eq:soft_thresh_pow}
\end{equation}
Similar form of connection function can be established for the third sub-class given by (\ref{eq:genrholom}-\ref{eq:genfdeltalom}) based on (\ref{eq:transformation}).
For all $\beta$ dependent connection functions defined above, at $\beta\rightarrow 0$ we obtain a linking kernel that becomes independent of the hidden variables, and thus, the generated network is an Erd{\H os}-R{\'e}nyi random graph. However, in the opposite case, when $\beta\rightarrow\infty$, the connection functions converge to the original 'hard' thresholded forms, and the generated networks are scale-free with a decay exponent of $\gamma=2$. Therefore, by tuning the parameter $\beta$ from $0$ to $\infty$ we can scan through a series of networks starting from the classical random graph, and arriving to a SF network obeying the inverse square decay law at the other end of the spectrum. Presumably, during this transition we may find a finite $\beta$ range in which the degree distribution is already power law like instead of the Poisson distribution, but the decay exponent $\gamma$ has not yet reached the $\gamma=2$ limit value.

\color{black}

\begin{figure}[htb!]
  \begin{center}
  \centering
    \includegraphics[width=\textwidth]{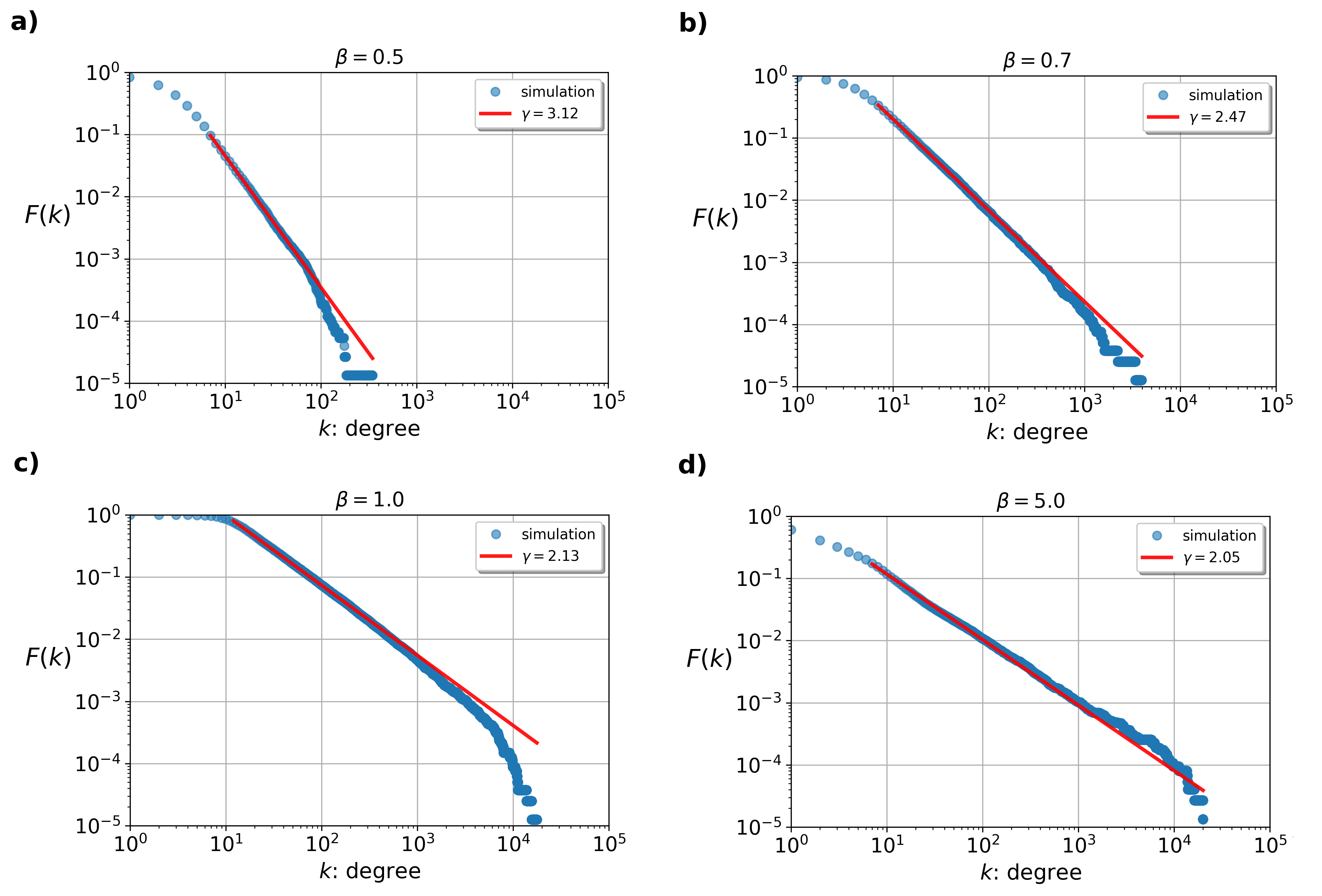}
    \caption{\small \color{black}{The complementary cumulative distribution of the node degrees $F(k)$ for four different networks, generated with the same $\rho(x)$ and $f(x,y)$, but with different $\beta$ parameters. The fitness distribution was chosen to be $\rho(x)=3x^2\exp(-x^3)$ and corresponding linking function $f(x,y)$ is given in (\ref{eq:soft_thresh_exp}). In panel a) we used $\beta=0.5$, and the decay characteristics of the resulting $F(k)$ seem to be close to that of SF networks with $\gamma\simeq 3.12$ (shown by the solid line). The parameter $\beta$ was increased to $\beta=0.7$ in panel b), where the decay of $F(k)$ suggests a $\gamma$ value of $\gamma\simeq 2.47$. In panels c) and d) we increased $\beta$ further to $\beta=1.0$ and $\beta=5.0$, reducing the $\gamma$ exponent to $\gamma=2.13$ and $\gamma=2.05$ respectively.In panel a) and b) $\Delta=\frac{1}{\beta}\ln N$}, while in panel c) and d) we used $\Delta=\ln N$ for obtaining sparse networks.} 
        \label{fig:beta}
  \end{center}
\end{figure}

Our simulation results shown in Fig.\ref{fig:beta}. provide a strong support for the assumption above. In the four panels we display the complementary cumulative degree distribution for an exponential-like model with $\rho(x)= 3x^2\mathrm{e}^{-x^3}$ at different $\beta$ parameters.

\subsection*{Characterizing the $\gamma(\beta)$ transition}
In this section we aim to characterize the main features of the above mentioned transition of $\gamma$ as a function of the effective temperature for the exponential case. In order to do so, we first define the general $\beta$ dependent integral for $k(x)$ and perform a transformation of variables $(z=H(x)+H(y)-\Delta)$:
\begin{equation}
    k(x)=N\int\limits_{0}^{\infty} \frac{H'(y) \mathrm{e}^{-H(y)} \mathrm{d}y}{1+\mathrm{e}^{-\beta(H(x)+H(y)-\Delta)}}
    =N\mathrm{e}^{H(x)-\Delta}\int\limits_{H(x)-\Delta}^{\infty} \frac{\mathrm{e}^{-z}}{1+\mathrm{e}^{-\beta z}}\mathrm{d}z=N\mathrm{e}^{H(x)-\Delta}F_{\beta,\Delta}(x)
\label{eq:beta_general_integral}
\end{equation}
offering a non-invertible form for the degree variable in general. However, approximate results can still be obtained possibly with logarithmic or sub-power corrections. If $H(x) \ll \Delta  $ and $\beta \in (0,1)$, the main contribution to the integral comes from the $z\ll 0$ range, where the value of the denominator is large. Based on that, the above expression can be approximated by
\begin{equation}
    k(x)=N\mathrm{e}^{H(x)-\Delta}\int\limits_{H(x)-\Delta}^{\infty} \frac{\mathrm{e}^{-z}}{1+\mathrm{e}^{-\beta z}}\mathrm{d}z \approx N\mathrm{e}^{H(x)-\Delta}\int\limits_{H(x)-\Delta}^{\infty} \mathrm{e}^{(\beta-1)z}\mathrm{d}z=\frac{N}{1-\beta}\mathrm{e}^{\beta(H(x)-\Delta)}.
    \label{eq:beta_general_approx}
\end{equation}
 Even though this approximation  tends to be less and less accurate as the value of $\beta$ converges to zero, its advantage is that it provides an analytic and invertible expression for the degree distribution written as
\begin{equation}
    p_{\beta}(k)\sim k^{-(1+\frac{1}{\beta})} + \textrm{sub-power corrections.}
\label{eq:beta_deg_dist}
\end{equation}
We note that similar forms have already been established in Refs. \cite{curvaturetemperaturecomplexnetw,Garlaschelli_Entropy}, however, not for the reversed Fermi-Dirac function. In addition to that, based on (\ref{eq:beta_general_approx}) we can also formulate an approximate condition for having a size independent degree distribution given by
\begin{equation}
    \Delta_{\beta}(N)\simeq\frac{1}{\beta} \ln N.
    \label{eq:indepsmallbeta}
\end{equation}
For $\beta > 1$ the integrand appearing in (\ref{eq:beta_general_integral}) becomes negligible for $z = H(x)-\Delta < 0$, hence the lower bound of the integral can approximately be replaced by zero as
\begin{equation}
    k(x)\approx N\mathrm{e}^{H(x)-\Delta}\int\limits_{0}^{\infty} \frac{\mathrm{e}^{-z}}{1+\mathrm{e}^{-\beta z}}\mathrm{d}z,
    \label{eq:beta_larger_than_one}
\end{equation}
which along similar arguments is yielding
\begin{equation}
    p_{\beta}(k)\sim k^{-2} + \textrm{sub-power corrections} \ \ \ \ \ \ \ \ \ \textrm{and}\ \ \ \ \ \ \ \ \  \Delta_{\beta}(N)\simeq \ln N.
    \label{eq:indeplargebeta}
\end{equation}
The results in (\ref{eq:indeplargebeta}) become exact in the limit of $\beta\to\infty$. Analogously, the above analysis applies to the power sub-class as well, where the average degree as a function of fitness is written as
\begin{equation}
    k(x)=N\int\limits_{1}^{\infty} \frac{G'(y) G^{-\alpha}(y)}{1+\left(\frac{G^{\alpha}(x)G^{\alpha}(y)}{\Delta}\right)^{-\beta}}\mathrm{d}y=
    NG^{\alpha-1}(x)\Delta^{-\alpha+1}\int\limits_{G(x)/\Delta}^{\infty} \frac{z^{-\alpha}}{1+z^{-\alpha\beta}}\mathrm{d}z,
\end{equation}
with $z=\frac{G(x)G(y)}{\Delta}$. Despite the same behaviour of the degree distribution, the formula above suggests that the condition for obtaining a size independent degree distribution requires 
$\Delta^{\alpha\beta}$ to be proportional to $N$ (for $\beta \in (0,1)$). Furthermore, it also implies that the accurate characterization of the $\gamma(\beta)$ transition for the three sub-classes requires similar considerations. 

\begin{figure}[htb!]
  \begin{center}
  \centering
    \includegraphics[width=0.5\textwidth]{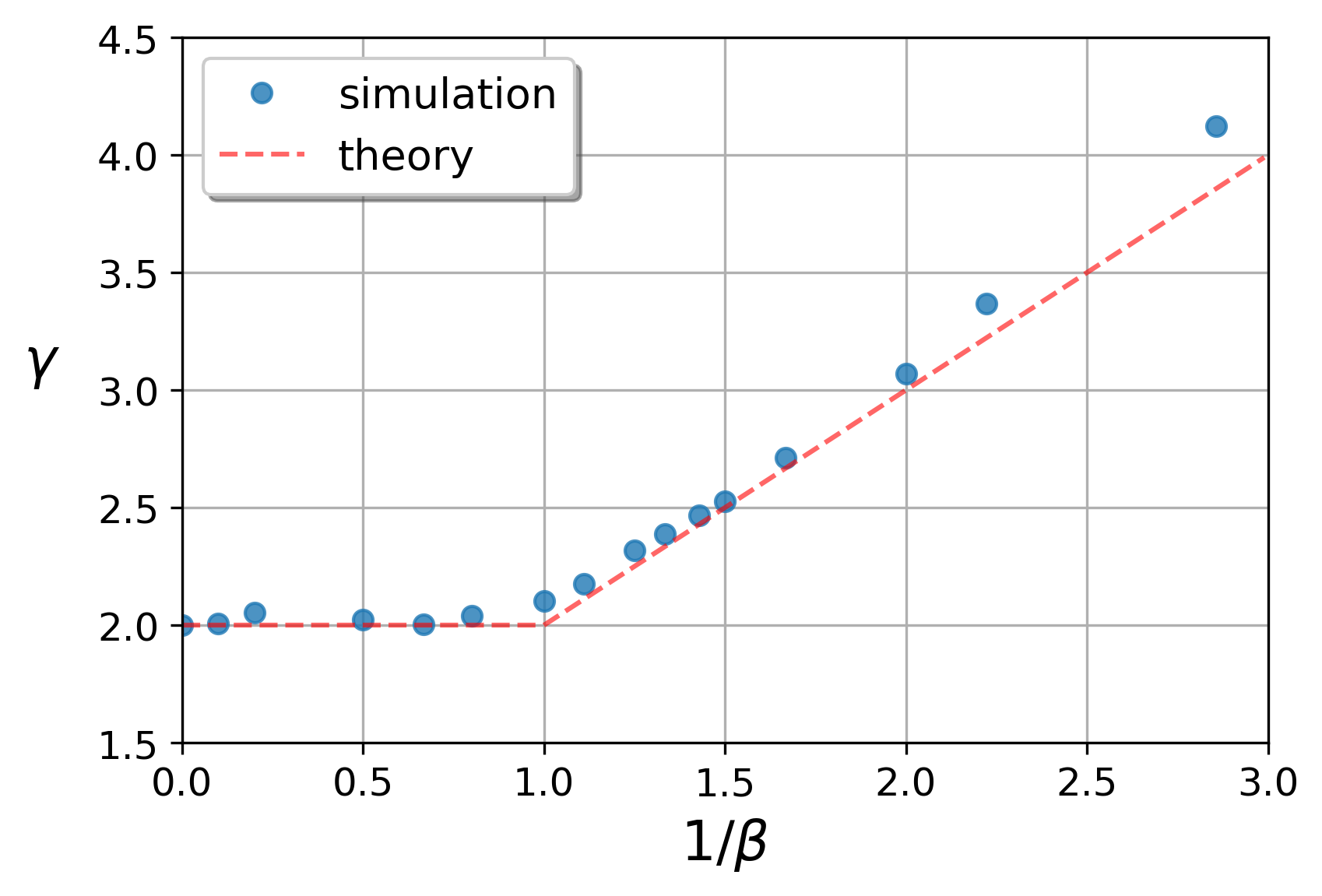}

      \caption{\color{black} Scaling exponent $\gamma$  of the degree distribution as a function of the effective temperature $1/\beta$ in the model with soft thresholding. The data points correspond to simulation results on networks of size  $N=20000$, where the fitness distribution was chosen to be $\rho(x)=3x^2\exp(-x^3)$ and the linking function $f(x,y)$ was given by (\ref{eq:soft_thresh_exp}). The dashed line shows the (approximate) analytic results given by (\ref{eq:beta_deg_dist}) and (\ref{eq:indeplargebeta}).}
      \label{fig:figtrans}
  \end{center}
\end{figure}

Our simulation results together with the approximation discussed above are shown in Fig.\ref{fig:figtrans}., depicting the transition of the scaling exponent as a function of the effective temperature $1/\beta$. We kept control over the average degree of the generated networks by relying on (\ref{eq:indepsmallbeta}) and (\ref{eq:indeplargebeta}), ensuring the size independence of the degree distribution.

\section*{Discussion}

We have revisited the inverse square decay law of non-geographical thresholded hidden variable models, which was already studied in Refs.\cite{caldarelli_first_peaked,masuda_inverse_square,fujihara_extreme_theory_hv} from different perspectives. We provided a far more general framework for thresholding linking mechanisms, where the form of the connection function $f(x,y)$ is not restricted to the usual Heaviside function, but instead can correspond to any general function with values in the $[0,1]$ interval, as long as $f(x,y)=0$ for a certain range of $x$ and $y$ values, and is non-zero outside this range. According to our results, this considerably weaker assumption on the form of the thresholded $f(x,y)$ allows a very broad range of connection functions, that combined with properly chosen fitness distributions $\rho(x)$ result in the inverse square decay law, similarly to the models discussed in Ref.\cite{caldarelli_first_peaked}.  Along this line we provided three general sub-classes of hidden variable distributions and accompanying connection functions (i.e., the exponential, the power-like and the mixed class) that all generate SF networks with a degree decay exponent of $\gamma=2$, and {\color{black} we also discussed how these different sub-classes are interrelated to each other. }
We also proposed a relaxation of the 'hard' threshold {\color{black} for each sub-class} imposed by the $\Delta$ controlled boundary in the connection functions. 
The basic idea was to use a 'reversed' Fermi-Dirac function providing a sigmoid transition between low and high linking probability values, where the sharpness of the transition (or in other words, the width of the intermediate linking probability values) is controlled by a parameter $\beta$. Based on analogy with temperature dependent graph ensembles \cite{Garlaschelli_Entropy}, $\beta$ can be interpreted as a sort of inverse temperature, where the original 'hard' thresholded models are recovered in the zero temperature limit of $\beta\rightarrow\infty$. The great advantage of the higher temperature models is that according to numerical simulations, the degree decay exponent becomes larger than $\gamma=2$, and by changing $\beta$, it can be tuned to any preferred value in the range of typical $\gamma$ values measured in real systems. {\color{black} We also generally discussed the criteria in multiple different cases of how to generate networks having degree distribution independent of the size.} Hence, the models with the relaxed threshold at finite $\beta$ values offer a flexible  fitness-based approach being adjustable to complicated fitness distributions  for generating {\color{black} sparse} SF networks with realistic degree decay exponent.

In conclusion, our analysis showed that linking kernels with a general lower-cutoff and having either additive, multiplicative or mixed dependence on their arguments can always generate SF networks together with the appropriately chosen fitness distributions. A further remarkable consequence of the above is that a general mapping can be established between different $\rho$ fitness distributions and possible $f$ linking functions. I.e., for any fitness distribution $\rho^{*}$ in general there exists a family of thresholded linking functions $f^{*}_{\Delta}$ that together give rise to scale-free networks with a $\gamma=2$
, and vice versa, for any thresholded linking function $f^{*}_{\Delta}$ we can find the corresponding fitness distribution $\rho^{*}$ together which they display the same property. This might provide an alternative way of understanding how those fitness/activity driven systems exhibit SF behaviour where the distributions of the hidden variables follow non-trivial, complicated forms.

\section*{Acknowledgements}
 This project has received funding from the European Union’s Horizon 2020 Research and Innovation Programme under Grant Agreement No. 740688 and was partially supported by the National Research, Development and Innovation Office under grant no. K128780.

\section*{Author contributions statement}

SGB, PP and GP developed the concept of the study, SGB derived the equations, SGB, PP, GP contributed to the interpretation of the results, SGB prepared the table and the figures, SGB, PP and GP wrote the paper.

\section*{Additional information}

\textbf{Competing interests:} The authors declare no competing interests. 

\end{document}